# MNTD: An Efficient Dynamic Community Detector Based on Nonnegative Tensor Decomposition

Hao Fang, Qu Wang, Qicong Hu, and Hao Wu

*Abstract*—Dynamic community detection is crucial for elucidating the temporal evolution of social structures, information dissemination, and interactive behaviors within complex networks. Nonnegative matrix factorization provides an efficient framework for identifying communities in static networks but fall short in depicting temporal variations in community affiliations. To solve this problem, this paper proposes a Modularity maximization-incorporated Nonnegative Tensor RESCAL Decomposition (MNTD) model for dynamic community detection. This method serves two primary functions: a) Nonnegative tensor RESCAL decomposition extracts latent community structures in different time slots, highlighting the persistence and transformation of communities; and b) Incorporating an initial community structure into the modularity maximization algorithm, facilitating more precise community segmentations. Comparative analysis of real-world datasets shows that the MNTD is superior to state-of-the-art dynamic community detection methods in the accuracy of community detection.

*Keywords*—Dynamic community detection, Tensor RESCAL decomposition, Modularity maximization, Temporal networks.

## I. INTRODUCTION

In today's interconnected world, complex networks are ubiquitous in all areas of human society, providing a powerful framework for modeling and understanding the structure and dynamics of diverse real-world systems, such as the Internet, social networks and biological networks. Therefore, it becomes an important tool for analyzing complex systems [1]-[7]. In the realm of network analysis, community detection stands as an essential technique for unraveling the intricate structures and functionalities inherent within complex networks. Through community detection, the network nodes are divided into several communities, where there are more connections within a community and fewer connections between communities. Community detection is widely studied in various fields due to its ability to reveal the structural organizations of a network and the functional relationships between different organizations. For instance, in neuroscience, community detection can uncover functional linkages leading to the identification of brain regions with similar functions [8], [9]. While in biology, it can discover functional units in protein-protein interaction networks [10]-[14].

Traditionally, nonnegative matrix factorization (NMF) is employed for static community detection [15]-[20], which has proven to be of high interpretability and scalability [21]-[26]. However, these methods perform poorly when capturing the dynamic evolution of networks, which is a critical aspect of understanding the evolving nature of real-world networks.

In fact, many networks are dynamic, meaning that their structures evolve over different time steps. Dynamic community detection aims to analyze evolving community structures. In social networks, the analysis of interaction patterns and the evolution of relationships between users helps to reveal how information spreads across communities and thus to adopt effective communication strategies for specific groups [27]-[32]. Similarly, the migrating of cancer cell communities is crucial for cancer diagnosis and treatment [33]-[35]. However, dynamic community detection presents new challenges, including efficiently incorporating temporal information and maintaining consistency across communities over different time frames. A promising approach is nonnegative tensor decomposition that extends the idea of nonnegative matrix factorization to a higher dimensional structure, which extracts community information from the decomposed low-rank components and tracks the evolution of the community over time [36]-[44].

In this study, we propose a Modularity maximization- incorporated Nonnegative Tensor RESCAL Decomposition (MNTD) model for dynamic community detection. The MNTD first adopts the nonnegative tensor RESCAL decomposition to represent a dynamic network, and then utilize the modularity maximization algorithm to achieve the desired community at each time slot. The main contributions of this work are summarized as follows:

1) The MNTD model is proposed which extends the static community discovery task to the dynamic network by nonnegative tensor RESCAL decomposition.
2) The input of modularity maximization algorithm is set as the community partition after tensor decomposition, and finally the further community structure is obtained.

In several real-world datasets, the MNTD model has advantages in most cases, and compared with the traditional modularity maximization algorithm, MNTD can obtain a more reasonable community structure.

The rest of this paper is organized as follow: Section II gives the related work. Section III gives the preliminaries. Section IV presents the proposed MNTD model. Section V reports the experimental results; and finally, Section VI concludes.

## II. Related Work

In this section, the relevant research on community detection is briefly reviewed.

### A. Static Community

In the past decade, community detection has emerged as a focal point of research, leading to the proposal of numerous outstanding models. Among them, NMF stands out for its high performance, having been recognized for its strong interpretability and scalability, attracting deep investigation by many researchers. Luo *et al.* [45] adjust the scale factor of the nonnegative multiplicative update through linear or nonlinear strategies to obtain a better community detector. He *et al.* [46] combine NMF with the auto encoder in graph neural networks to improve the nonlinear representation capability of the model. Liu *et al.* [47] use multiple latent factor matrices to represent large-scale undirected networks, and apply regularization constraints to these matrices to enhance representation learning. However, these algorithms perform community discovery tasks in static networks without considering the dynamic development of the network.

### B. Dynamic Community

In order to make the traditional matrix-based model can deal with large-scale dynamic network better. Ma *et al.* [48] propose an evolutionary nonnegative matrix factorization method that introduces a time-smoothing term into the objective function to capture the changes of communities over time. However, the computational cost of NMF for each adjacency matrix becomes prohibitive with an increase in the number of time slices. Laetitia *et al.* [49] extract the community activity structure of temporal networks through nonnegative tensor CP factorization. Esraa *et al.* [8] use the Tucker decomposition to determine the subspace that best describes the community structure of brain partitions. However, CP factorization decomposes the data into a linear combination of factors, so it is difficult to intuitively reveal the interaction patterns between nodes and the dynamic changes in community structure [50]-[54]. Additionally Tucker factorization is not specifically designed to capture relationships between nodes, which makes it poor at detecting community dynamics [55]-[59].

### C. Modularity Maximization

Modularity is a crucial metric for assessing the quality of community divisions, reflecting the difference between the actual edge density within the community and the expected density in random scenarios [60]-[63]. So, researchers propose some modularity maximization algorithms. Newman *et al.* [64] propose a greedy algorithm Fast Newman (FN) which find each local optimal value and finally integrates the local optimal value into the approximate optimal value of the whole. Blondel *et al.* [65] propose Louvain algorithm, it enhances the modularity by quickly merging communities based on local modularity improvements.

However, modularity maximization algorithms share certain drawbacks, such as a heavy dependence on the initial community partitioning, which is often random and can lead to suboptimal local maxima. Additionally, these algorithms focus on modularity improvement, which might inadvertently change the number of communities to enhance the modularity score, leading to irrational community structures.

In summary, the tensor decomposition method proposed in this paper overcomes the shortcoming of traditional static community detection algorithm which cannot capture the dynamic characteristics of a network. This method uses tensor decomposition to obtain the initial community, and uses it as the input of modularization maximization algorithm to derive further refined community structure, which can alleviate the inherent limitations of modularization maximization algorithm to a certain extent.

## III. Preliminaries

### A. Notations

Given a set of undirected temporal network $\mathbf{G}=(G^1, G^2,\ldots G^{(t)})$ with node set at each time $v^{(t)}=(v_1,\ldots v_n^{(t)})$ where $t$ is the total number of time slices and $n^{(t)}$ is the total number of nodes at time $t$. For the network under each time slice, corresponding adjacency matrices are defined as $\mathbf{W}=(W^1, W^2,\ldots W^{(t)})$, the element $w_{ij}^{(t)}$ denote whether there is an edge between $v_i$ and $v_j$ in $G^{(t)}$, if in time $t$ there is an edge $w_{ij}^{(t)}=1$ and vice versa, it is 0. We first assume $k$ is the number of communities; $N$ is the maximum number of nodes in the dynamic network. The community indicative matrices sequence is defined as $\mathbf{B}=(B^1, B^2,\ldots B^{(t)})$. Each indicative matrix $B^t \in \mathbb{R}^{N \times k}$, element $b_{ij}^t$ represents the probability that node $i$ belongs to community $j$ at time $t$. Finally, $C=(c^1, c^2,\ldots c^{(t)})$ is defined as a community membership list, where the element $c^{(t)}$ in the list is a one-dimensional array containing the community to which each node belongs at time $t$.

### B. Tensor RESCAL Decomposition

As shown in Fig.1. RESCAL factorizes a multi-relation tensor $\mathbf{X} \in \mathbb{R}^{N \times N \times T}$ into a matrix $A \in \mathbb{R}^{N \times k}$ and a tensor $\mathbf{R} \in \mathbb{R}^{k \times k \times T}$ and the transpose of A. Through this decomposition, low-dimensional representations of entities can be extracted.

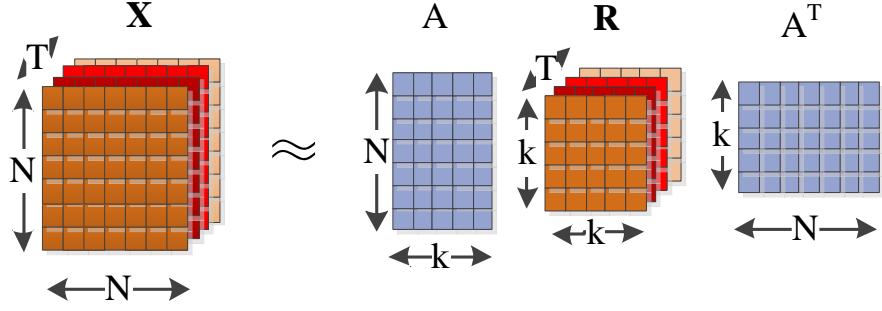

Fig. 1. Tensor RESCAL Decomposition.

Compared with other tensor decomposition methods, RESCAL decomposition pays more attention to the information at each slice. In this study, the core idea of RESCAL decomposition is to represent each frontal slice of the third-order tensor as the product of the matrix A which representing the latent features of the nodes and the matrix $R_t$ which describing the interactions mode between the latent features of the nodes. Therefore, given a tensor composed of adjacency matrices at different times, the community membership of different nodes at different times can be analyzed by RESCAL decomposition. The form of RESCAL decomposition can be represented as [66]-[68]:

$$X_t \approx AR_t A^T, \quad \text{for } t=1,2,\ldots T. \qquad (1)$$

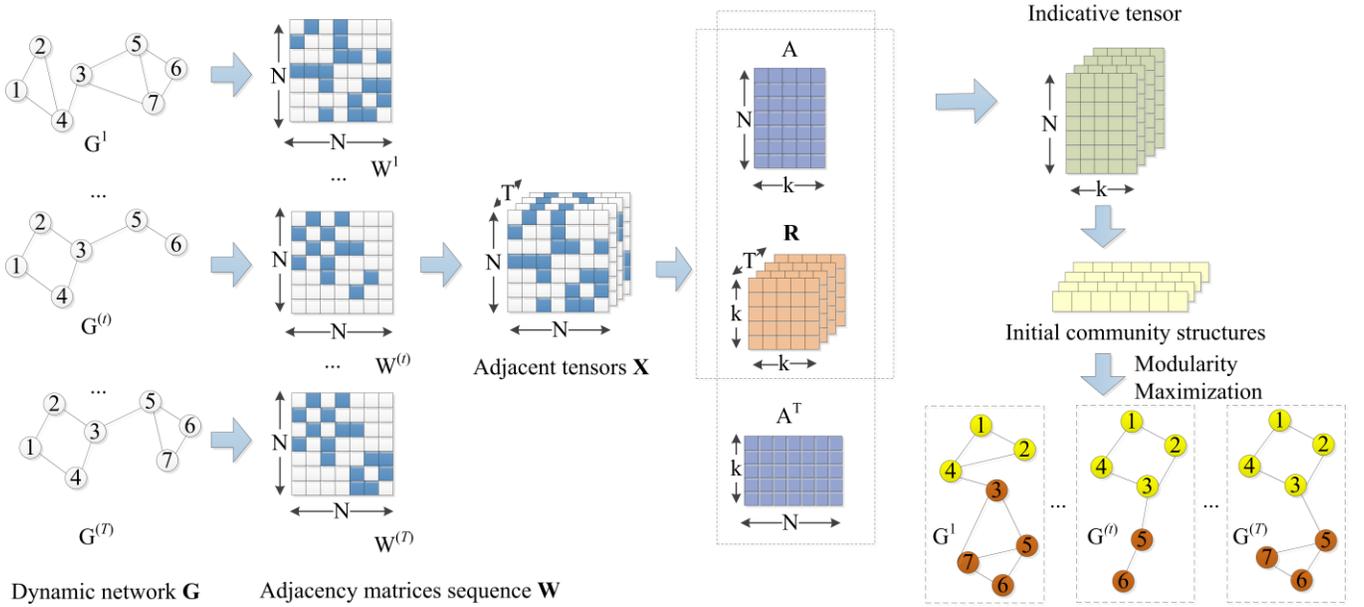

Fig.2. Framework of MNTD model

## IV. THE MNTD MODEL

In this section, we denote the solution objective, give the nonnegative update rule of the method, and finally analyze the complexity of the algorithm. The working details of the MNTD model are shown in Fig.2. A dynamic network is divided into several time slices, each of which builds an adjacency matrix. These matrices are then stacked to form a adjacency tensor $\mathbf{X} \in \mathbb{R}^{N \times N \times T}$. Multiply each slice of the factor matrix A and tensor R obtained by RESCAL decomposition to get the community indication tensor, thus obtaining the initial community membership of each node. The further community structure is obtained by modularity maximization algorithm.

### A. Learning Objectives and Updating Scheme

For the tensor $\mathbf{X} \in \mathbb{R}^{N \times N \times T}$, its low-rank approximation is constructed using matrix A and tensor $\mathbf{R}$. The objective function is formed using the Frobenius norm and a regularization term [69]-[75]:

$$L = \frac{1}{2}\sum_t \| X_t - AR_t A^T \|_F^2 + \frac{1}{2}\left( \lambda_A \| A \|_F^2 + \lambda_R \sum_t \| R_t \|_F^2 \right), \qquad (2)$$

where $\|\cdot\|_F^2$ is Frobenius norm, $\lambda_A$ and $\lambda_R$ are regularization coefficients.

It is unreasonable for nodes to have negative community membership values. Therefore, it is necessary to constrain the matrix A and the tensor **R** to be nonnegative:

$$L = \frac{1}{2}\sum_t \| X_t - AR_t A^T \|_F^2 + \frac{1}{2}\left( \lambda_A \| A \|_F^2 + \lambda_R \sum_t \| R_t \|_F^2 \right),$$
$$s.t.\ A \geq 0, R_t \geq 0,\ t=\{1,2,\ldots T\}. \quad (3)$$

Next, least square is used to update A and **R**, finally achieving a multiplicative update of the nonnegative RESCAL decomposition [76]-[84]:

$$A \leftarrow A \cdot \frac{\sum_t X_t AR_t^T + X_t^T AR_t}{A\left(\left[\sum_t R_t A^T AR_t^T + R_t^T A^T AR_t\right] + \lambda_A I\right)}. \quad (4)$$

$$R_t \leftarrow R_t \cdot \frac{A^T X_t A}{A^T AR_t A^T A + \lambda_R R_t}. \quad (5)$$

*B. Detection of Initial Community*

After obtaining the final matrix A, and the matrix $R_t$ at each time point, the community indicator matrices are derived by matrix multiplication:

$$B^t = AR_t,\ \text{for}\ t=1,2,\ldots T. \quad (6)$$

The indicative matrix $B^t$ delineates the network's soft community memberships at time point $t$ [85]-[89]. These matrices are collectively referred to as List **B**, and then the community members of each node at each time point are obtained by:

$$c_i^{(t)} = \arg\max\left(b_{i.}^t\right),\quad \text{for}\ t=1,2,\ldots T. \quad (7)$$

where $b_{i.}^t$ is the $i$-th row in matrix $B^t$. Note that the timing change of the network will inevitably lead to the disappearance of nodes. Therefore, to obtain a more accurate community structure, nodes whose rows and columns are 0 in the adjacency matrix are removed from the community index at the corresponding time.

*C. Modularity Maximization Enhances Community*

In order to identify a better community structure, after obtaining the initial communities through RESCAL decomposition, a modularity maximization module is employed to enhance the community structure. We repeat the last two steps of the Louvain algorithm until the modularity no longer increases. a) For each node, attempt to move it into the community where its neighbors reside. If such a move increases the modularity, perform it. Repeat this process for all nodes until it is no longer possible to further increase modularity. b) Aggregating existing communities into super nodes to form a new reduced network.

The above is the specific method used in our model. We summarize our approach in Algorithm 1.

| Algorithm 1 MNTD for dynamic community detection |
|---|
| Input: Dynamic network **G**, parameters $\lambda_A$, $\lambda_R$, maximum number of iterations $\varphi$, number of communities $k$.<br>Output: List of community members for each time segment C=($c^1$, $c^2$,…,$c^{(t)}$).<br>1. Set number of iterations $n$=0, list C=∅. Randomly initialize the matrix A and the tensor **R**.<br>2. **while** $n<\varphi$ and not converged **do**<br>3.    **Update** A by (4)<br>4.    **Update R** by (5)<br>5.    $n=n+1$<br>6. **end while**<br>7. The initial community structure C is obtained by (6) (7).<br>8. **Update** community structure C by the modularity maximization algorithm. |

*D. Time Complexity Analysis*

Here, we analyze the time complexity of Algorithm 1. Because the rank of RESCAL decomposition is much smaller than the number of nodes $N$, the time complexity of A and $R_t$ optimization process is $O(N^2r)$, where $r$ is the rank of RESCAL decomposition. Obtaining the initial community requires multiplying each slice of **R** by A, which is $O(N^2rt)$. Finally, the time complexity of the Louvain algorithm in each time slice is $O(N^2t)$ at worst, but tensor decomposition can provide an excellent initial community partition, which can significantly reduce the computation time.

## V. EXPERIMENTS

In this section, we introduce some dynamic network datasets, evaluation metrics for community detection, and select some state-of-the-art models for comparison.

*A. Datasets*

We choose four real-world datasets and the details of them are shown in Table I. PrimaryD1 and PrimaryD2 [45] are datasets with community ground truth, while chess [90] and cellphone [91] are datasets without ground truth.
1) ***PrimaryD1 and PrimaryD2:*** the two datasets record the interactions of students in a primary school over two days, respectively. We divide the first day into 7 time slices and the second day into 8 time slices by the hours.
2) ***Chess:*** chess is a dataset that records chess player data from 1998 to 2006, we define the game of two players as an edge.
3) ***Cellphone:*** cellphone is a dataset containing 10 days of phone exchanges of *Paraiso* members in June 2006, we record the call between two callers as an edge.

TABLE I
DYNAMIC NETWORK DATA SET STATISTICS

| Datasets | Nodes | Edges | Times | Communities |
|---|---|---|---|---|
| **D1:PrimaryD1** | 242 | 12290 | 7 | 13 |
| **D2:PrimaryD2** | 242 | 12741 | 8 | 13 |
| **D3:Cellphone** | 400 | 512 | 10 | / |
| **D4:Chess** | 7301 | 66833 | 9 years | / |

*B. Tested Models*

We compare MNTD to state-of-the-art models for dynamic community detection: TMOGA [92], Cr-ENMF [48], DECS [93], and ECD [94] to verify its effectiveness. Besides we chose a model using the nonnegative tensor CP decomposition (NCPD) [49] to verify that RESCAL decomposition provides excellent initial community membership, we also add modularity-enhanced module to this model (MENCPD).

Ablation experiments that remove the modularity maximization module of MNTD and use only modularity maximization to optimize the random initial community structure are also set up. For the coefficients of these models, we use the default value, and for MNTD we set $\lambda_A = 0.2$, $\lambda_R = 0.07$.

*C. Evaluation Metrics*

   *1) Modularity*

Higher modularity means the connections among the vertices of the same community are denser, while the connections among the vertices of different communities are sparser. For a temporal network its modularity at time $t$ is defined as follows:

$$Q(t) = \frac{1}{2L^{(t)}} \sum_{ij} \left[ w_{ij}^{(t)} - \frac{d_i^{(t)} d_j^{(t)}}{2L^{(t)}} \right] \delta\left(c_i^{(t)}, c_j^{(t)}\right),$$

where $d_i^{(t)}$ is the degree of the $v_i$ at time $t$, $L^{(t)}$ is the sum of the weights of all edges in the network at time $t$, $\delta(c_i^{(t)}, c_j^{(t)})$ is an indicator function, if at the time $t$, $v_i$ and $v_j$ belong to the same community, then $\delta(c_i^{(t)}, c_j^{(t)})=1$, otherwise, it is 0.

   *2) Normalized Mutual Information (NMI)*

Consistent with [95]-[97], we use NMI to evaluate the performance of MNTD. NMI measures the degrees of information sharing between actual communities.

$$\text{NMI}\left(U^{(t)}, V^{(t)}\right) = \frac{-2\sum_{i=1}^{C_{U^{(t)}}} \sum_{j=1}^{C_{V^{(t)}}} N_{ij}^{(t)} \log\left(\frac{N_{ij}^{(t)} N^{(t)}}{N_i^{(t)} N_j^{(t)}}\right)}{\sum_{i=1}^{C_{U^{(t)}}} N_i^{(t)} \log\left(\frac{N_i^{(t)}}{N^{(t)}}\right) + \sum_{j=1}^{C_{V^{(t)}}} N_j^{(t)} \log\left(\frac{N_j^{(t)}}{N^{(t)}}\right)},$$

where $U^{(t)}$ and $V^{(t)}$ are sets of nodes in the actual community and the community detected by the clustering algorithm in the time slice $t$, respectively. $N_{ij}^{(t)}$ is the number of nodes that belong to both $U^{(t)}$ and $V^{(t)}$ in the time slice $t$. $C_U^{(t)}$ and $C_V^{(t)}$ are the number of communities in the node set $U^{(t)}$ and $V^{(t)}$ in the time slice $t$, respectively. The larger the NMI, the results are closer to the actual community.

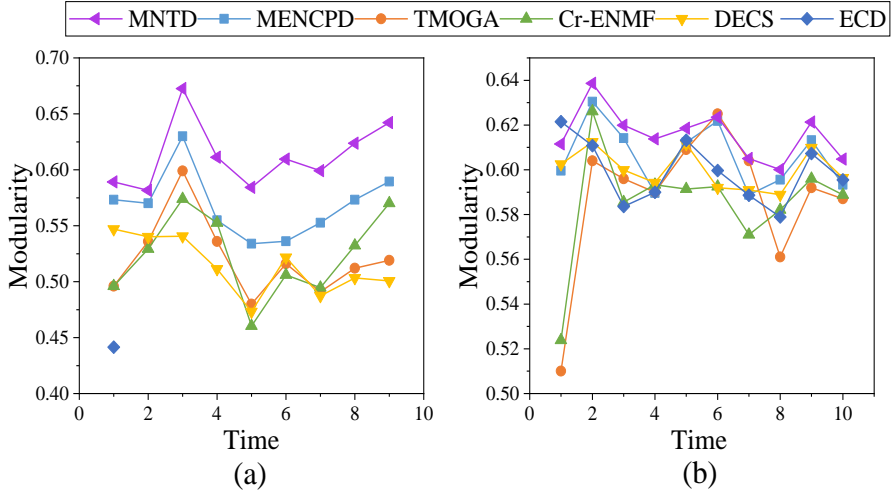

Fig. 3 The average modularity of different models in each time slice of two datasets without ground truth. (a) Chess and (b) Cellphone.

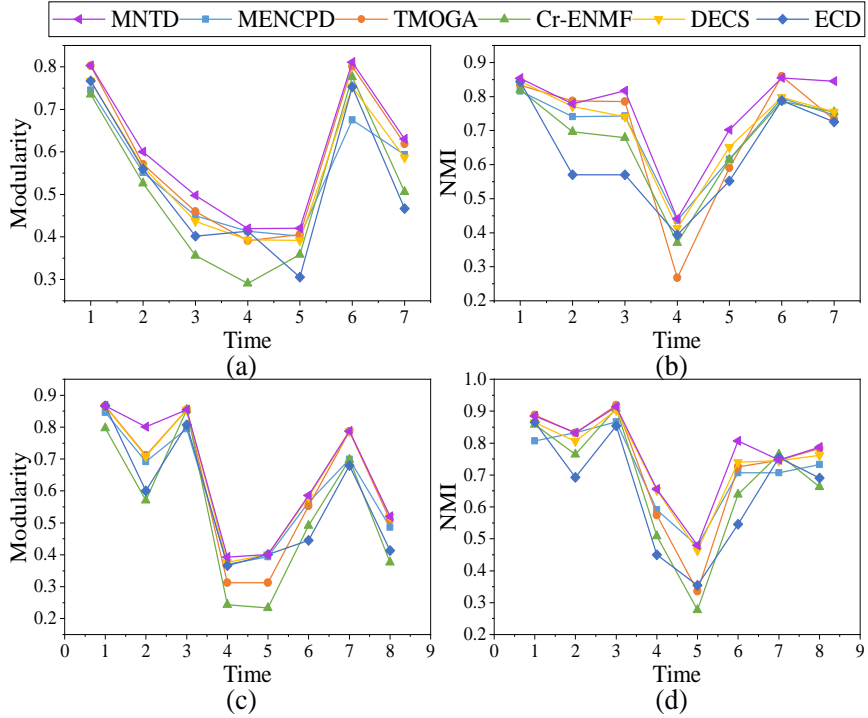

Fig. 4 The average modularity and NMI in each time slice of two datasets with ground truth. (a)-(b) are PrimaryD1. (c)-(d) are PrimaryD2.

TABLE II
NMI (MEAN+STD) ON PRIMARYD1 AND PRIMARYD2

| Models<br>Datasets | MNTD | TMOGA | Cr-ENMF | DECS | ECD |
|---|---|---|---|---|---|
| D1 | **0.756±0.14** | 0.693±0.21 | 0.674±0.15 | 0.711±0.14 | 0.635±0.16 |
| D2 | **0.763±0.14** | 0.725±0.20 | 0.673±0.20 | 0.742±0.14 | 0.651±0.18 |

### D. Dynamic Community Detection Results

We compare the MNTD to state-of-the-art models, and after analyzing the results, we get the following founds:

*a*) **MNTD's performance in detecting community structures is superior to its peers.** To test the performance of each model on datasets without ground truth, each method is assessed 20 times on each dataset, resulting in an averaged modularity score. Fig. 3 presents the average modularity scores for various models across different time points for both the mobile phone and chess datasets. It clearly indicates that MNTD significantly enhances the accuracy of community detection. This improvement is due to the decomposition of co-learning node relationships across time using RESCAL, and the enhancement of community structure by modularity maximization algorithms In particular, as can be seen in Fig. 3(a), the ECD model can only identify the community of time 1 in the chess dataset, which reflects its inability to depict the dynamic information of large networks.

*b*) **In most cases, the community structure identified by MNTD is closer to the real community structure than its peers.** Because MNTD uses the modularity maximization algorithm, it is one-sided to evaluate community partitions only from modularity. To provide a more comprehensive experimental result, further experiments are conducted on two datasets with ground truth. Besides modularity, NMI is employed to assess the quality of community detection. As can be seen in Fig. 4, the MNTD has advantages in most cases. Although TMOGA and DECS get slightly better results at some time points, their performance are characterized by significant fluctuations. For example, their results at time 4 on primaryD1 and time 5 on primaryD2 is very bad. Table II displays the average NMI results for the various models. MNTD perform 8.31% and 4.98% better than TMOGA on NMI metrics on both datasets; 5.9% and 2.75% better than DECS.

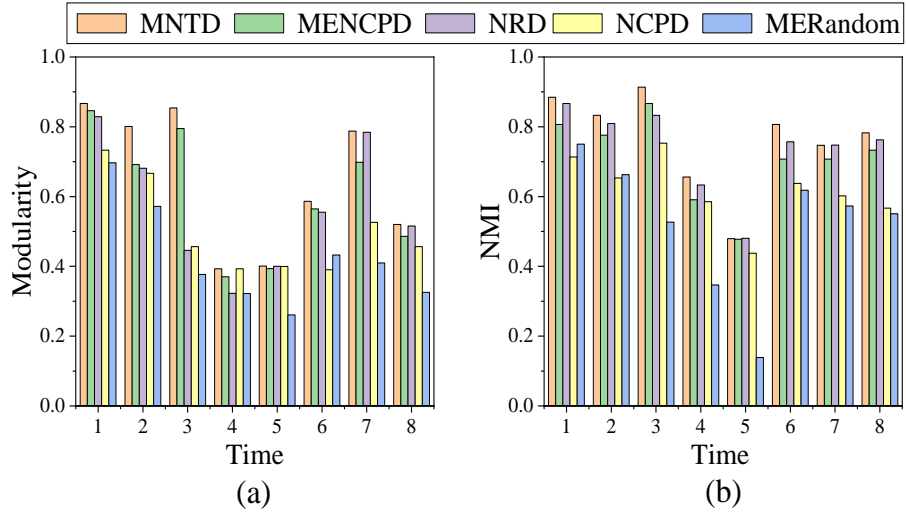

Fig. 5 Ablation experiments on PrimaryD2. (a) and (b) show the modularity and NMI of the community structure obtained from several models, respectively.

TABLE III
THE NMI (MEAN+STD) OF ABLATION EXPERIMENTS ON
PRIMARYD1 AND PRIMARYD2

| Models<br>Datasets | MNTD | MENCPD | NRD | NCPD | MERandom |
|---|---|---|---|---|---|
| **D1** | **0.756±0.14** | 0.710±0.13 | 0.711±0.14 | 0.683±0.10 | 0.711±0.20 |
| **D2** | **0.763±0.14** | 0.715±0.12 | 0.722±0.13 | 0.619±0.10 | 0.520±0.19 |

*c*) **Tensor decomposition can provide an excellent community structure, and RESCAl decomposition performs better than CP decomposition.** Notably, MNTD performs significantly better than MENCPD, another model based on tensor decomposition and modularized maximization strategies. So we continue to explore the performance of these two methods without modularity reinforcement. An ablation study compares the quality of community division using only RESCAL and CP decomposition. In addition, the effect of modularity-enhance community membership in random initialization (MERandom) is also studied. The results in Fig.5 show that compared with CP decomposition, RESCAL decomposition provides a more accurate community structure, and the community structure obtained by MNTD is superior to the model of modularity maximization based on random initialization. This improvement can be attributed to the fact that RESCAL can model the time relationship of nodes more accurately than CP through co-learning multiple time relationship matrices.

*d*) **Starting from an excellent initial community structure can reduce the limitations of the modularity maximization algorithm and improve the accuracy of community detection.** Although MNTD improves modularity, the NMI obtained by the model is sometimes slightly lower than the NRD; for example at time 5 and time 7 of the PrimaryD2 data, NRD is 2.12% and 0.08% better than MNTD respectively. The summary of the measurement results for the models mentioned in the ablation experiment is presented in Table III. In the PrimaryD2 dataset, MNTD outperforms the NRD model by 5.37% and MERandom by 31.8%. This also demonstrates that providing a random initialization for community partitioning leads to poor locally optimal solutions, whereas MNTD can avoid this problem. On average, from the perspective of high-quality community structure, the community structure will not be unreasonable because the modularity maximization algorithm only focuses on modularity. Instead, in most cases, it optimizes the initial high-quality community structure.

## VI. Conclusion

In this study, a community detection model MNTD is proposed. The excellent initial community structure obtained from the decomposition of the nonnegative tensor RESCAL is provided as input to the modularity maximization algorithm. Compared with several state-of-the-art models, the accuracy of MNTD is superior to them. The ablation experiments prove that MNTD can reduce the limitations brought by the modularity maximization algorithm, and also shows that community structure provided by RESCAL decomposition is effective. In future, we plan to combine graph neural network (GNN) with tensor factorization to enhance the detection effect of time-evolving network community dynamics.

## Reference

[1] X. Luo, H. Wu, Z. Wang, J. Wang, and D. Meng, "A Novel Approach to Large-Scale Dynamically Weighted Directed Network Representation," *IEEE Trans. Pattern Analysis and Machine Intelligence*, vol. 44, no. 12, pp. 9756-9773, 2022.

[2] W. Qin, X. Luo, and M. Zhou, "Adaptively-Accelerated Parallel Stochastic Gradient Descent for High-Dimensional and Incomplete Data Representation Learning," *IEEE Trans. Big Data*, vol. 10, no. 1, pp. 92-107, 2024.

[3] T. He, Y. Liu, Y.-S. Ong, X. Wu, and X. Luo, "Polarized message-passing in graph neural networks," *Artificial Intelligence*, vol. 331, p. 104129, 2024.

[4] M. Habibi and P. Khosravi, "Disruption of Protein Complexes from Weighted Complex Networks," *IEEE/ACM Trans. Computational Biology and Bioinformatics*, vol. 17, no. 1, pp. 102-109, 2020.

[5] D. Wu, X. Luo, M. Shang, Y. He, G. Wang, and X. Wu, "A Data-Characteristic-Aware Latent Factor Model for Web Services QoS Prediction," *IEEE Trans. Knowledge and Data Engineering*, vol. 34, no. 6, pp. 2525-2538, 2022.

[6] X. Luo, Y Zhou, Z Liu, and MC Zhou, "Fast and accurate non-negative latent factor analysis of high-dimensional and sparse matrices in recommender systems," *IEEE Trans. Knowledge and Data Engineering*, vol. 35, no. 4, pp. 3897-3911, 2021.

[7] H. Wu, Y. Xia and X. Luo, "Proportional-Integral-Derivative-Incorporated Latent Factorization of Tensors for Large-Scale Dynamic Network Analysis," *2021 China Automation Congress (CAC)*, Beijing, China, 2021, pp. 2980-2984.

[8] E. Al-sharoa, M. Al-khassaweneh, and S. Aviyente, "Tensor Based Temporal and Multilayer Community Detection for Studying Brain Dynamics During Resting State fMRI," *IEEE Trans. Biomedical Engineering,* vol. 66, no. 3, pp. 695-709, 2019.

[9] Z. Lin and H. Wu, "Dynamical Representation Learning for Ethereum Transaction Network via Non-negative Adaptive Latent Factorization of Tensors," *2021 International Conference on Cyber-Physical Social Intelligence (ICCSI)*, Beijing, China, 2021, pp. 1-6.

[10] L. Hu, S. Yang, X. Luo, H. Yuan, K. Sedraoui, and M. Zhou, "A Distributed Framework for Large-scale Protein-protein Interaction Data Analysis and Prediction Using MapReduce," *IEEE/CAA Journal of Automatica Sinica*, vol. 9, no. 1, pp. 160-172, 2022.

[11] X. Luo, Y. Zhou, Z. Liu, L. Hu, and M. Zhou, "Generalized Nesterov's Acceleration-Incorporated, Non-Negative and Adaptive Latent Factor Analysis," *IEEE Trans. Services Computing*, vol. 15, no. 5, pp. 2809-2823, 2022.

[12] Z. Tang, X. Kou, H. Feng, D. Cui, W. Jia, and L. Li, "Semi-Supervised Protein-Protein Interactions Extraction Method Based on Knowledge Distillation and Virtual Adversarial Training," *in 2022 IEEE International Conference on Bioinformatics and Biomedicine (BIBM)*, 2022, pp. 3887-3889.

[13] X. Luo, L. Wang, P. Hu, and L. Hu, "Predicting Protein-Protein Interactions Using Sequence and Network Information via Variational Graph Autoencoder," *IEEE-ACM Trans. Computational Biology and Bioinformatics*, vol. 20, no. 5, pp. 3182-3194, 2023.

[14] A. Purohit, S. Acharya, and J. Green, "A novel Greedy approach for Sequence based Computational prediction of Binding-Sites in Protein-Protein Interaction," in *2021 IEEE 21st International Conference on Bioinformatics and Bioengineering (BIBE)*, 2021, pp. 1-8.

[15] H. Zhou, T. He, Y.-S. Ong, G. Cong, and Q. Chen, "Differentiable Clustering for Graph Attention," *IEEE Trans. Knowledge and Data Engineering*, vol. 36, no. 8, pp. 3751-3764, 2024.

[16] X. Luo, Y. Zhou, Z. Liu, and M. Zhou, "Fast and Accurate Non-Negative Latent Factor Analysis of High-Dimensional and Sparse Matrices in Recommender Systems," *IEEE Trans. Knowledge and Data Engineering*, vol. 35, no. 4, pp. 3897-3911, 2023.

[17] L. Hu, Y. Yang, Z. Tang, Y. He, and X. Luo, "FCAN-MOPSO: An Improved Fuzzy-Based Graph Clustering Algorithm for Complex Networks With Multiobjective Particle Swarm Optimization," *IEEE Trans. Fuzzy Systems*, vol. 31, no. 10, pp. 3470-3484, 2023.

[18] B. Wu, X. Yao, and B. Zhang, "Semi-supervised Community Detection using Graph Embedding," in *2022 IEEE International Conference on e-Business Engineering (ICEBE)*, 2022, pp. 141-147..

[19] D. Cheng, J. Huang, S. Zhang, X. Zhang, and X. Luo, "A Novel Approximate Spectral Clustering Algorithm With Dense Cores and Density Peaks," *IEEE Trans. Systems, Man, and Cybernetics: Systems*, vol. 52, no. 4, pp. 2348-2360, 2022.

[20] C. Tran, W.-Y. Shin, and A. Spitz, "Community Detection in Partially Observable Social Networks," *ACM Trans. Knowl. Discov. Data*, vol. 16, no. 2, pp. 1-24, 2022.

[21] H. Wu, X. Luo, and M. Zhou, "Neural Latent Factorization of Tensors for Dynamically Weighted Directed Networks Analysis," *2021 IEEE International Conference on Systems, Man, and Cybernetics (SMC)*, 2021, pp. 3061-3066.

[22] Z. Liu, Y. Yi, and X. Luo, "A High-Order Proximity-Incorporated Nonnegative Matrix Factorization-Based Community Detector," *IEEE Trans. Emerging Topics in Computational Intelligence*, vol. 7, no. 3, pp. 700-714, 2023.

[23] M. Ortiz-Bouza and S. Aviyente, "Orthogonal Nonnegative Matrix Tri-Factorization for Community Detection in Multiplex Networks," *in ICASSP 2022 IEEE International Conference on Acoustics, Speech and Signal Processing (ICASSP)*, 2022, pp. 5987-5991.

[24] J. Wang, W. Li, and X. Luo, "A distributed adaptive second-order latent factor analysis model", *IEEE/CAA Journal of Automatica Sinica*, pp. 1-3, 2024.

[25] X. Luo, Y. Zhong, Z. Wang, and M. Li, "An Alternating-Direction-Method of Multipliers-Incorporated Approach to Symmetric Non-Negative Latent Factor Analysis," *IEEE Trans. Neural Networks and Learning Systems*, vol. 34, no. 8, pp. 4826-4840, 2023.

[26] D. Kamuhanda, M. Wang, and K. He, "Sparse Nonnegative Matrix Factorization for Multiple-Local-Community Detection," *IEEE Trans. Computational Social Systems*, vol. 7, no. 5, pp. 1220-1233, 2020.

[27] W Li, X Luo, H Yuan, MC Zhou, "A momentum-accelerated Hessian-vector-based latent factor analysis model," *IEEE Trans. Services Computing*, vol. 16, no. 2, pp. 830-844, 2022.

[28] T. He, Y. Ong, and L. Bai, "Learning Conjoint Attentions for Graph Neural Nets," in *Neural Information Processing Systems*, 2021.

[29] Q. Wang, X. Liu, T. Shang, Z. Liu, H. Yang, and X. Luo, "Multi-Constrained Embedding for Accurate Community Detection on Undirected Networks," *IEEE Trans. Network Science and Engineering*, vol. 9, no. 5, pp. 3675-3690, 2022.

[30] D. Wu, Y. He, and X. Luo, "A Graph-Incorporated Latent Factor Analysis Model for High-Dimensional and Sparse Data," *IEEE Trans. Emerging Topics in Computing*, vol. 11, no. 4, pp. 907-917, 2023.

[31] R. Xu, Q. Zhang, and S. Tan, "The Formation of Reciprocal Social Support in Online Support Groups: A Network Modeling Approach," *IEEE Trans. Computational Social Systems*, vol. 10, no. 6, pp. 3370-3384, 2023.

[32] F. Bi, X. Luo, B. Shen, H. Dong, and Z. Wang, "Proximal alternating-direction-method-of-multipliers-incorporated nonnegative latent factor analysis," *IEEE/CAA Journal of Automatica Sinica*, vol. 10, no. 6, pp. 1388-1406, 2023.

[33] T. C. G. A. Network, "Oncogenic Signaling Pathways in The Cancer Genome Atlas," *Cell*, vol.173, no. 2, pp. 321-337, 2018.

[34] H. U. Osmanbeyoglu, R. Pelossof, J. F. Bromberg, and C. S. Leslie, "Linking signaling pathways to transcriptional programs in breast cancer," *Genome research*, vol. 24, no. 11, p. 1869-1880, 2014.

[35] R. Al-Bahrani, A. Agrawal, and A. Choudhary, "Colon cancer survival prediction using ensemble data mining on SEER data," in *2013 IEEE International Conference on Big Data*, 2013, pp. 9-16.

[36] H. Wu, X. Luo, and M. Zhou, "Advancing Non-Negative Latent Factorization of Tensors With Diversified Regularization Schemes," *IEEE Trans. Services Computing*, vol. 15, no. 3, pp. 1334-1344, 2022.

[37] X. Luo, H. Wu, and Z. Li, "Neulft: A Novel Approach to Nonlinear Canonical Polyadic Decomposition on High-Dimensional Incomplete Tensors," *IEEE Trans. Knowledge and Data Engineering*, vol. 35, no. 6, pp. 6148-6166, 2023.

[38] C.-Y. Wang, Y.-L. Gao, J.-X. Liu, X.-Z. Kong, and C.-H. Zheng, "Single-Cell RNA Sequencing Data Clustering by Low-Rank Subspace Ensemble Framework," *IEEE/ACM Trans. Computational Biology and Bioinformatics*, vol. 19, no. 2, pp. 1154-1164, 2022.

[39] F. Bi, T. He, Y. Xie, and X. Luo, "Two-Stream Graph Convolutional Network-Incorporated Latent Feature Analysis," *IEEE Trans. Services Computing*, vol. 16, no. 4, pp. 3027-3042, 2023.

[40] C. Lyu, Y. Shi, and L. Sun, "A Novel Local Community Detection Method Using Evolutionary Computation," *IEEE Trans. Cybernetics*, vol. 51, no. 6, pp. 3348-3360, 2021.

[41] H. Wu, X. Luo, M. Zhou, M. J. Rawa, K. Sedraoui, and A. Albeshri, "A PID-incorporated Latent Factorization of Tensors Approach to Dynamically Weighted Directed Network Analysis," *IEEE/CAA Journal of Automatica Sinica*, vol. 9, no. 3, pp. 533-546, 2022.

[42] F. Bi, T. He, and X. Luo, "A Fast Nonnegative Autoencoder-Based Approach to Latent Feature Analysis on High-Dimensional and Incomplete Data," *IEEE Trans. Services Computing*, vol. 17, no. 3, pp. 733-746, 2024.

[43] C. Lyu, Y. Shi, L. Sun, and C.-T. Lin, "Community Detection in Multiplex Networks Based on Evolutionary Multitask Optimization and Evolutionary Clustering Ensemble," *IEEE Trans. Evolutionary Computation*, vol. 27, no. 3, pp. 728-742, 2023.

[44] M. Chen, C. He, and X. Luo, "MNL: A Highly-Efficient Model for Large-scale Dynamic Weighted Directed Network Representation," *IEEE Trans. Big Data*, vol. 9, no. 3, pp. 889-903, 2023.

[45] X. Luo, Z. Liu, L. Jin, Y. Zhou, and M. Zhou, "Symmetric Nonnegative Matrix Factorization-Based Community Detection Models and Their Convergence Analysis," *IEEE Trans. Neural Networks and Learning Systems*, vol. 33, no. 3, pp. 1203-1215, 2022.

[46] C. He, Y. Zheng, X. Fei, H. Li, Z. Hu, and Y. Tang, "Boosting Nonnegative Matrix Factorization Based Community Detection With Graph Attention Auto-Encoder," *IEEE Trans. Big Data*, vol. 8, no. 4, pp. 968-981, 2022.

[47] Z. Liu, X. Luo, and M. Zhou, "Symmetry and Graph Bi-Regularized Non-Negative Matrix Factorization for Precise Community Detection," *IEEE Trans. Automation Science and Engineering*, pp. 1-15, 2024.

[48] X. Ma, B. Zhang, C. Ma, and Z. Ma, "Co-regularized nonnegative matrix factorization for evolving community detection in dynamic networks," *Information Sciences*, vol. 528, pp. 265-279, 2020.

[49] L. Gauvin, A. Panisson, and C. Cattuto, "Detecting the Community Structure and Activity Patterns of Temporal Networks: A Non-Negative Tensor Factorization Approach," *Plos One*, vol. 9, no. 1, p. e86028, 2014.

[50] Y. Yuan, R. Wang, G. Yuan, and L. Xin, "An Adaptive Divergence-Based Non-Negative Latent Factor Model," *IEEE Trans. Systems, Man, and Cybernetics: Systems*, vol. 53, no. 10, pp. 6475-6487, 2023.

[51] D. Wu, Q. He, X. Luo, M. Shang, Y. He, and G. Wang, "A Posterior-Neighborhood-Regularized Latent Factor Model for Highly Accurate Web Service QoS Prediction," *IEEE Trans. Services Computing*, vol. 15, no. 2, pp. 793-805, 2022.

[52] E. Gujral, R. Pasricha, and E. Papalexakis, "Beyond Rank-1: Discovering Rich Community Structure in Multi-Aspect Graphs," in *Proceedings of The Web Conference 2020*, 2020, pp. 452-462.

[53] W. Qin, H. Wang, F. Zhang, J. Wang, X. Luo, and T. Huang, "Low-Rank High-Order Tensor Completion With Applications in Visual Data," *IEEE Trans. Image Processing*, vol. 31, pp. 2433-2448, 2022.

[54] X. Luo, H. Wu, H. Yuan, and M. Zhou, "Temporal Pattern-Aware QoS Prediction via Biased Non-Negative Latent Factorization of Tensors," *IEEE Trans. Cybernetics*, vol. 50, no. 5, pp. 1798-1809, 2020.

[55] X. Luo, Y Yuan, S Chen, N Zeng, and Z Wang, "Position-transitional particle swarm optimization-incorporated latent factor analysis," *IEEE Trans. Knowledge and Data Engineering*, vol. 34, no. 8, pp 3958-3970,2022.

[56] Y.-D. Kim and S. Choi, "Nonnegative Tucker Decomposition," in *2007 IEEE Conference on Computer Vision and Pattern Recognition*, 2007, pp. 1-8.

[57] H. Wu and X. Luo, "Instance-Frequency-Weighted Regularized, Nonnegative and Adaptive Latent Factorization of Tensors for Dynamic QoS Analysis," *2021 IEEE International Conference on Web Services*, 2021, pp. 560-568.

[58] X. Shi, Q. He, X. Luo, Y. Bai, and M. Shang, "Large-Scale and Scalable Latent Factor Analysis via Distributed Alternative Stochastic Gradient Descent for Recommender Systems," *IEEE Trans. Big Data*, vol. 8, no. 2, pp. 420-431, 2022.

[59] M. Chen, R. Wang, Y. Qiao, and X. Luo, "A Generalized Nesterov's Accelerated Gradient-Incorporated Non-Negative Latent-Factorization-of-Tensors Model for Efficient Representation to Dynamic QoS Data," *IEEE Trans. Emerging Topics in Computational Intelligence*, vol. 8, no. 3, pp. 2386-2400, 2024.

[60] M. E. J. Newman, "Fast algorithm for detecting community structure in networks," *Physical Review E*, vol. 69, no. 6, p. 066133, 2004.

[61] A. Bartal and G. Ravid, "Member Behavior in Dynamic Online Communities: Role Affiliation Frequency Model," *IEEE Trans. Knowledge and Data Engineering*, pp. 1-1, 2020.

[62] J. Cheng, M. Chen, M. Zhou, S. Gao, C. Liu, and C. Liu, "Overlapping Community Change-Point Detection in an Evolving Network," *IEEE Trans. Big Data*, vol. 6, no. 1, pp. 189-200, 2020.

[63] J. Xiao, Y.-C. Zou, and X.-K. Xu, "A Metaheuristic-Based Modularity Optimization Algorithm Driven by Edge Directionality for Directed Networks," *IEEE Trans. Network Science and Engineering*, pp. 1-14, 2023.

[64] M. E. J. Newman and M. Girvan, "Finding and Evaluating Community Structure in Networks," *Physical Review E*, vol. 69, no. 2, p. 026113, 2004.

[65] V. D. Blondel, J. L. Guillaume, R. Lambiotte, and E. Lefebvre, "Fast unfolding of communities in large networks," *Journal Of Statistical Mechanics-theory And Experiment*, vol. 2008, no. 10, p. 10008, 2008.

[66] M. Nickel, V. Tresp, and H.-P. Kriegel, "Factorizing YAGO: scalable machine learning for linked data," *Proceedings of the 21st international conference on World Wide Web*, 2012, pp. 271-280.

[67] D. Wu, P. Zhang, Y. He, and X. Luo, "MMLF: Multi-Metric Latent Feature Analysis for High-Dimensional and Incomplete Data," *IEEE Trans. Services Computing*, vol. 17, no. 2, pp. 575-588, 2024.

[68] M. Bhattarai et al., "Distributed non-negative RESCAL with automatic model selection for exascale data," *Journal of Parallel and Distributed Computing*, vol. 179, p. 104709, 2023.

[69] Y. Yuan, Q. He, X. Luo, and M. Shang, "A multilayered-and-randomized latent factor model for high-dimensional and sparse matrices," *IEEE trans. Big Data*, vol. 8, no. 3, pp. 784-794, 2020.

[70] B. Bader, R. A. Harshman, and T. G. Kolda, "Temporal Analysis of Semantic Graphs Using ASALSAN," in *Seventh IEEE International Conference on Data Mining (ICDM 2007)*, 2007, pp. 33-42.

[71] H. Wu, X. Luo, and M. Zhou, "Advancing Non-Negative Latent Factorization of Tensors With Diversified Regularization Schemes," *IEEE Trans. Services Computing*, vol. 15, no. 3, pp. 1334-1344, 2022.

[72] J. Guan, B. Chen, and X. Huang, "Community Detection via Autoencoder-Like Nonnegative Tensor Decomposition," *IEEE Trans. Neural Networks and Learning Systems*, vol. 35, no. 3, pp. 4179-4191, 2024.

[73] X. Luo, J. Chen, Y. Yuan, and Z. Wang, "Pseudo Gradient-Adjusted Particle Swarm Optimization for Accurate Adaptive Latent Factor Analysis," *IEEE Trans. Systems, Man, and Cybernetics*, vol. 54, no. 4, pp. 2213-2226, 2024.

[74] D Wu, X Luo, Y He, and M Zhou "A prediction-sampling-based multilayer-structured latent factor model for accurate representation to high-dimensional and sparse data," *IEEE Trans. Neural Networks and Learning Systems*, vol. 35, no. 3, pp. 3845-3858, 2022.

[75] M. Nickel, V. Tresp, and H. P. Kriegel, "A Three-Way Model for Collective Learning on Multi-Relational Data," in *International Conference on International Conference on Machine Learning*, 2011.

[76] Z. Li, S. Li, O. O. Bamasag, A. Alhothali, and X. Luo, "Diversified Regularization Enhanced Training for Effective Manipulator Calibration," *IEEE Trans. Neural Networks and Learning Systems*, vol. 34, no. 11, pp. 8778-8790, 2023.

[77] X. Chen, X. Luo, L. Jin, S. Li, and M. Liu, "Growing Echo State Network With an Inverse-Free Weight Update Strategy", *IEEE Trans. Cybernetics*, vol. 53, no. 2, pp. 753-764, 2023.

[78] Y. Qin, G. Feng, Y. Ren, and X. Zhang, "Block-Diagonal Guided Symmetric Nonnegative Matrix Factorization," *IEEE Trans. Knowledge and Data Engineering*, vol. 35, no. 3, pp. 2313-2325, 2023.

[79] Y. Song, M. Li, X. Luo, G. Yang, and C. Wang, "Improved Symmetric and Nonnegative Matrix Factorization Models for Undirected, Sparse and Large-Scaled Networks: A Triple Factorization-Based Approach," *IEEE Trans. Industrial Informatics*, vol. 16, no. 5, pp. 3006-3017, 2020.

[80] F. Bi, T. He, and X. Luo, "A Two-Stream Light Graph Convolution Network-based Latent Factor Model for Accurate Cloud Service QoS Estimation," in *2022 IEEE International Conference on Data Mining (ICDM)*, 2022, pp. 855-860.

[81] D. Krompaß, M. Nickel, X. Jiang, and V. Tresp, "Non-negative tensor factorization with rescal," *ECMUPKDD 2013 Workshop on Tensor Methods for Machine Learning*, 2013, pp.1-10.

[82] J. Wang, N. Guan, X. Huang, and Z. Luo, "Constraint-Relaxation Approach for Nonnegative Matrix Factorization: A Case Study," in *2015 IEEE International Conference on Systems, Man, and Cybernetics*, 2015, pp. 2192-2197.

[83] D. Wu, Z. Li, Z. Yu, Y. He, and X. Luo, "Robust Low-Rank Latent Feature Analysis for Spatiotemporal Signal Recovery," *IEEE Trans. Neural Networks and Learning Systems*, pp. 1-14, 2023.

[84] X. Xu, M. Lin, X. Luo, and Z. Xu, "HRST-LR: A Hessian Regularization Spatio-Temporal Low Rank Algorithm for Traffic Data Imputation," *IEEE Trans. Intelligent Transportation Systems*, vol. 24, no. 10, pp. 11001-11017, 2023.

[85] P. Jiao, T. Li, H. Wu, C.-D. Wang, D. He, and W. Wang, "HB-DSBM: Modeling the Dynamic Complex Networks From Community Level to Node Level," *IEEE Trans. Neural Networks and Learning Systems*, vol. 34, no. 11, pp. 8310-8323, 2023.

[86] Z. Wang et al., "Large-Scale Affine Matrix Rank Minimization With a Novel Nonconvex Regularizer," *IEEE Trans. Neural Networks and Learning Systems*, vol. 33, no. 9, pp. 4661-4675, 2022.

[87] A. Gorovits, E. Gujral, E. E. Papalexakis, and P. Bogdanov, "LARC: Learning Activity-Regularized Overlapping Communities Across Time," in *Proceedings of the 24th ACM SIGKDD International Conference on Knowledge Discovery & Data Mining*, 2018, pp. 1465-1474.

[88] Y. Song, M. Li, Z. Zhu, G. Yang, and X. Luo, "Nonnegative Latent Factor Analysis-Incorporated and Feature-Weighted Fuzzy Double c-Means Clustering for Incomplete Data," *IEEE Trans. Fuzzy Systems*, vol. 30, no. 10, pp. 4165-4176, 2022.

[89] T. Li et al., "Exploring Temporal Community Structure via Network Embedding," *IEEE Trans. Cybernetics*, vol. 53, no. 11, pp. 7021-7033, 2023.

[90] H. Qin, R.-H. Li, G. Wang, X. Huang, Y. Yuan, and J. X. Yu, "Mining Stable Communities in Temporal Networks by Density-Based Clustering," *IEEE Trans. Big Data*, vol. 8, no. 3, pp. 671-684, 2022.

[91] F. Folino and C. Pizzuti, "An evolutionary multiobjective approach for community discovery in dynamic networks," *IEEE Trans. Knowledge and Data Engineering*, vol. 26, no. 8, pp. 1838-1852, 2014.

[92] J. Zou, F. Lin, S. Gao, G. Deng, W. Zeng, and G. Alterovitz, "Transfer Learning Based Multi-Objective Genetic Algorithm for Dynamic Community Detection," *arXiv preprint,* arXiv.2109.15136. 2021.

[93] F. Liu, J. Wu, S. Xue, C. Zhou, J. Yang, and Q. Sheng, "Detecting the evolving community structure in dynamic social networks," *World Wide Web*, vol. 23, no. 2, pp. 715-733, 2020.

[94] F. Liu, J. Wu, C. Zhou and J. Yang, "Evolutionary Community Detection in Dynamic Social Networks," *International Joint Conference on Neural Networks (IJCNN)*, 2019, pp. 1-7.

[95] L. Hu, X. Pan, Z. Tang, and X. Luo, "A Fast Fuzzy Clustering Algorithm for Complex Networks via a Generalized Momentum Method," *IEEE Trans. Fuzzy Systems*, vol. 30, no. 9, pp. 3473-3485, 2022.

[96] X. Meng, S. Wang, Z. Liang, D. Yao, J. Zhou, and Y. Zhang, "Semi-supervised anomaly detection in dynamic communication networks," *Information Sciences*, vol. 571, pp. 527-542, 2021.

[97] W. Luo, B. Duan, H. Jiang, and L. Ni, "Time-Evolving Social Network Generator Based on Modularity: TESNG-M," *IEEE Trans. Computational Social Systems*, vol. 7, no. 3, pp. 610-620, 2020.